\documentclass{raa}           
\usepackage{graphicx,times}
\usepackage{natbib}
\usepackage{amssymb,amsmath}
\bibpunct{(}{)}{;}{a}{}{,}

\usepackage[a4paper=true,pagebackref=true]{hyperref} 
\hypersetup{colorlinks = true, linkcolor = blue, anchorcolor = red, citecolor = 
blue, filecolor = red, pagecolor = red, urlcolor = red}

\begin{document}

   \title{Inner jet kinematics and the viewing angle towards the $\gamma$-ray 
narrow-line Seyfert\,1 galaxy 1H\,0323+342
}

 \volnopage{ {\bf 2016} Vol.\ {\bf X} No. {\bf XX}, 000--000}
   \setcounter{page}{1}

   \author{
L. Fuhrmann,\inst{1}\thanks{E-mail: fuhrmann@zess.uni-siegen.de}
V. Karamanavis,\inst{1}\thanks{E-mail: vkaramanavis@mpifr-bonn.mpg.de}
S. Komossa,\inst{1}\thanks{E-mail: astrokomossa@gmx.de}
E. Angelakis,\inst{1}
T. P. Krichbaum,\inst{1}
R. Schulz,\inst{2,3}
A. Kreikenbohm,\inst{2,3}
M. Kadler,\inst{2}
I. Myserlis,\inst{1}
E. Ros,\inst{1,4,5}
I. Nestoras,\inst{1}
J. A. Zensus\inst{1}
   }

   \institute{
Max-Planck-Institut f\"ur Radioastronomie, Auf dem H\"ugel 69, D-53121 
Bonn, Germany\\
\and
Lehrstuhl f\"ur Astronomie, Universit\"at W\"urzburg, Campus Hubland Nord,
Emil-Fischer-Stra\ss e 31, D-97074 W\"urzburg, Germany\\
\and
Dr. Karl Remeis-Sternwarte \& Erlangen Centre for Astroparticle Physics,
Sternwartstr. 7, D-96049 Bamberg, Germany\\
\and
Observatori Astron\`{o}mic, Universitat de Val\`{e}ncia, C/
Catedr\'{a}tico Jos\'{e} Beltr\'{a}n 2, 46980 Paterna, Val\`{e}ncia, Spain\\
\and
Departament d'Astronomia i Astrof\'{i}sica, Universitat de Val\`{e}ncia,
C/ Dr. Moliner 50, 46100 Burjassot, València, Spain\\
\vs \no
   {\small Received XXXX; accepted Aug. 2016}
}

\abstract{
Near-Eddington accretion rates onto low-mass black holes are thought
 to be a prime driver of the multi-wavelength properties of
  the narrow-line Seyfert 1 (NLS1) population
  of active galactic nuclei (AGN). Orientation effects have repeatedly
  been considered as another important factor involved, but detailed
  studies have been hampered by the lack of measured viewing angles
  towards this type of AGN. 
   Here we present multi-epoch, 15\,GHz VLBA images (MOJAVE program) of the 
radio-loud and {\it Fermi}/LAT-detected NLS1 galaxy
  1H\,0323+342. These are combined with
  single-dish, multi-frequency radio monitoring of the source's
  variability, obtained with the Effelsberg 100-m and IRAM 30-m
  telescopes, in the course of the F-GAMMA program. The VLBA images reveal 6
components with apparent speeds of $\sim$1--7\,c, and one quasi-stationary
feature.
Combining the obtained apparent jet speed ($\beta_{{\rm app}}$) and variability
Doppler factor ($D_{\rm var}$) estimates together with other methods, we
constrain the viewing angle $\theta$ towards 1H\,0323+342 to 
  $\theta \leq 4^{\circ}$--$13^{\circ}$. Using literature values of
  $\beta_{{\rm app}}$ and $D_{\rm var}$, we also deduce a viewing angle
  of ${\le} 8^{\circ}$--$9^{\circ}$ towards another radio- and
  $\gamma$-ray-loud NLS1, namely SBS 0846+513. 
}

   \authorrunning{L. Fuhrmann, V. Karamanavis, S. Komossa et al. }            
   \titlerunning{Jet kinematics and the viewing angle towards the $\gamma$-NLS1 

galaxy 1H\,0323+342}  
   \maketitle

%
\section{Introduction}           
auto-capitalized
Narrow-line Seyfert 1 (NLS1) galaxies are a subclass of active
 galactic nuclei (AGNs) with extreme multi-wavelength 
properties. They
 have been defined as AGN with small widths of their broad Balmer lines
 (Full width at half maximum, FWHM(H$\beta$) < 2000\,km/s) and weak
[OIII]5007/H$\beta$ emission
 \citep[e.g.,][see review by
\citealt{2008RMxAC..32...86K}]{1985ApJ..297...166,
 1992ApJS...80..109B, 2004AJ....127.1799G, 2012AJ....143...83X}. As a
class, they are characterized by super-strong FeII emission
 complexes \citep{2001A&A...372..730V, 2004AJ....127.1799G,
2006ApJS..166..128Z}, rapid X-ray
 variability and soft X-ray spectra \citep{1999ApJS..125..297L,
 1996A&A...305...53B, 2010ApJS..187...64G}, strong outflows
\citep{2008ApJ...680..926K, 2015A&A...574A.121K},  
host galaxies with a preference
 for pseudo bulges \citep{2011MNRAS.417.2721O, 2012ApJ...754..146M},
 and a deficiency of {\em very} radio-loud systems, even though
 7\% of the sources of the whole population are radio-loud 
\citep{2006AJ....132..531K}.
 A few of the radio-loud systems have been detected in $\gamma$-rays
by the Large Area Telescope (LAT) onboard the \textit{Fermi} Gamma-ray Space
Telescope for the first time \citep[e.g.][]{2009ApJ...707L.142A,
2012MNRAS.426..317D, 2015MNRAS.454L..16Y}. The majority of these 
are characterized by one-sided radio jets, implying that we do not see
these systems edge-on 
\cite[e.g.][]{2009ApJ...707..727A, 2011ApJ...738..126D, 2012MNRAS.426..317D,
2013MNRAS.436..191D,
2014ApJ...781...75W, 2015MNRAS.453.4037O, 2015arXiv151102631S} .

 Several lines of evidence suggest that near-Eddington accretion rates
 onto low-mass black holes are a prime driver of the multi-waveband 
 appearance of the NLS1 population,
 which, however, cannot account for all of the NLS1 properties 
\citep{2008RMxAC..32...86K}.
 Source orientation with respect to the observer
 has repeatedly been considered as another important
 factor. Arguments in favor of, or against, the importance of
 orientation effects in NLS1 galaxies have been given, based on
 emission-line widths, line profiles, polarimetry, radio properties and
 other considerations \citep[e.g.,][]{1985ApJ..297...166,
1992MNRAS.256..589P, 2000ApJ...536L...5S, 2005MNRAS.359..846S,
2006A&A...456...75C, 2006AJ....132..531K, 2011arXiv1109.4181P,
2013MNRAS.431..210C}.
 Firm conclusions have been hampered so far by the lack of a method of
 actually {\em measuring} the viewing angle towards NLS1 galaxies.
 In particular, the knowledge of orientation is of great importance to 
measure the black hole massess of NLS1 galaxies (and 
other AGNs) accurately, when applying scaling relations 
\citep[][]{2011arXiv1109.4181P} in case the
 broad-line region (BLR) is flattened \citep[][]{2006MNRAS.369..182J,
 2011arXiv1109.4181P}.

The focus of this paper is a viewing-angle measurement
of the radio-loud NLS1 galaxy 1H\,0323+342.
 This galaxy, at redshift $z$=0.0629, was recognized as a radio-loud NLS1
 galaxy, sharing the properties of NLS1 galaxies
 and blazars by \cite{2007ApJ...658L..13Z}. Its host galaxy displays a 
peculiar 
morphology in form of a one-armed spiral \citep{2007ApJ...658L..13Z} or ring 
structure
 \citep{2008A&A...490..583A, 2014ApJ...795...58L}. Several independent
estimates imply a black hole mass of $10^7$ solar masses
 and an
accretion rate near the Eddington rate 
\citep[e.g.][]{2007ApJ...658L..13Z, 
2015AJ....150...23Y}.
 1H\,0323+342 is highly variable at all wavebands including the
 radio \citep{2014ApJ...781...75W, 2015A&A...575A..55A},
optical \citep{2014PASJ...66..108I}, X-rays \citep{2009AdSpR..43..889F,
2014ApJ...789..143P, 2015AJ....150...23Y} 
and $\gamma$ rays
 \citep{2009ApJ...707L.142A, 2011MNRAS.413.2365C, 2013ATel.5344....1C,
2015AJ....149...41P}.
It is one of only a few radio-loud NLS1 galaxies detected at GeV energies by 
the
{\it Fermi Gamma-ray Space Telescope} \citep{2009ApJ...707L.142A}. Monitoring
and spectroscopy with {\it Swift} and {\it Suzaku}, along with SED modelling,
suggest that the UV and X-ray emission is dominated by the 
accretion disk and
corona \citep[][see also \citealp{2015ApJ...798...43S}]{2015AJ....150...23Y}.
Radio observations of 1H\,0323+342 revealed a flat radio spectrum
\citep{1994A&AS..106..303N, 2012ApJ...760...41D, 2015A&A...575A..55A},
high apparent brightness temperatures
\citep{2014ApJ...781...75W, 2015A&A...575A..55A}, and a core with extended
structure of 15\,kpc observed with the VLA \citep{2008A&A...490..583A}.
At 8\,GHz, \cite{2014ApJ...781...75W} reported the presence of a compact
core-jet structure at parsec (pc) scales and evidence for two slowly moving jet
components ($\beta_{\rm app} \ll$ c).

Aiming to constrain the viewing angle towards 
the source,  we present multi-epoch, high-resolution VLBI 
images of 1H\,0323+342 at 15\,GHz \citep[see][for first 
results]{2015PhDT.......232K} as
well as single-dish radio monitoring  observations obtained with the Effelsberg
100-m and IRAM 30-m  telescopes during the years 2010--2014. 
 The former are used to measure the parsec-scale
 jet structure and its kinematics over a time span of 3 years, and the latter 
to
infer the brightness temperature and Doppler factor for the source. Their
combination provides estimates of the viewing angle towards this NLS1 galaxy.
The paper is structured as  follows. 
Section \ref{sect:2} describes the observations and
data analysis  procedures. In Sections \ref{D_var} and \ref{kinematics}, we 
present our results from the total flux density and VLBI monitoring, 
respectively. Section \ref{counter_jet_ratio} concerns the estimation of the 
jet-to-counter-jet ratio for 1H\,0323+342, and Sect. \ref{sect:Combi} presents 
the calculation of its viewing angle combining our single-dish and VLBI 
findings, along with a similar estimate for  SBS\,0846+513.
We provide a discussion of our results in Sect. \ref{sect:Disc}, and a short 
summary with conclusions in Sect. \ref{sect:summary}. Throughout this paper we 
adopt the following cosmological parameters:
 ${\rm H}_{0} = 71\,{\rm km\,s}^{-1}\,{\rm Mpc}^{-1}$, $\Omega_{\rm m} = 0.27$
 and $\Omega_{\Lambda} = 0.73$ \citep[][]{2007ApJS..170..377S}. At the redshift
of 1H\,0323+342 an angular separation of 1\,milliarcsecond (mas) corresponds to
1.196\,pc and a proper motion of 1\,mas\,yr$^{-1}$ translates to an apparent
superluminal speed of 3.89\,c. We also follow the convention $S \propto
\nu^{+\alpha}$, where $S$ is the radio flux density, $\nu$ the observing
frequency and $\alpha$ the optically thin spectral index. 

\section{Observations and data analysis}\label{sect:2}

\subsection{Single-dish, total flux-density monitoring}

Since July 2010, the NLS1 galaxy 1H\,0323+342 has been regularly
observed at cm to short-mm bands in the framework of a dedicated
monitoring program of NLS1 galaxies at the Effelsberg (EB) 100-m telescope
\citep[see][]{2015A&A...575A..55A} and in the course of the {\it Fermi}-related
F-GAMMA\footnote{\url{www.mpifr-bonn.mpg.de/div/vlbi/fgamma/fgamma.html}}
monitoring program \citep{2007AIPC..921..249F, 2010arXiv1006.5610A,
2014MNRAS.441.1899F} including quasi-simultaneous
observations with the IRAM 30-m telescope (at Pico Veleta, PV). These
observations are closely coordinated with 
the more general AGN flux density monitoring conducted at the IRAM 30-m
telescope \citep[e.g.][]{1998ASPC..144..149U}.
The overall frequency range of the 1H\,0323+342 monitoring data spans from 2.64
to 142\,GHz. 
The EB measurements were conducted with cross-scans  
in the frequency range from
2.64 to 43.05\,GHz. 
The PV observations were carried out with calibrated cross-scans using the 
Eight
MIxer Receiver (EMIR) at 86.2 and 142.3 GHz. 
In the data reduction process for each station, pointing offset,
elevation-dependent gain, atmospheric opacity and sensitivity corrections have
been applied to the data \citep[for more details, see][Nestoras et al., 
submitted]{2008A&A...490.1019F, 2014MNRAS.441.1899F, 2009A&A...501..801A}.

\begin{figure}
\centering
\includegraphics[width=0.5\linewidth,trim={0mm 0 0
0.001mm},clip]{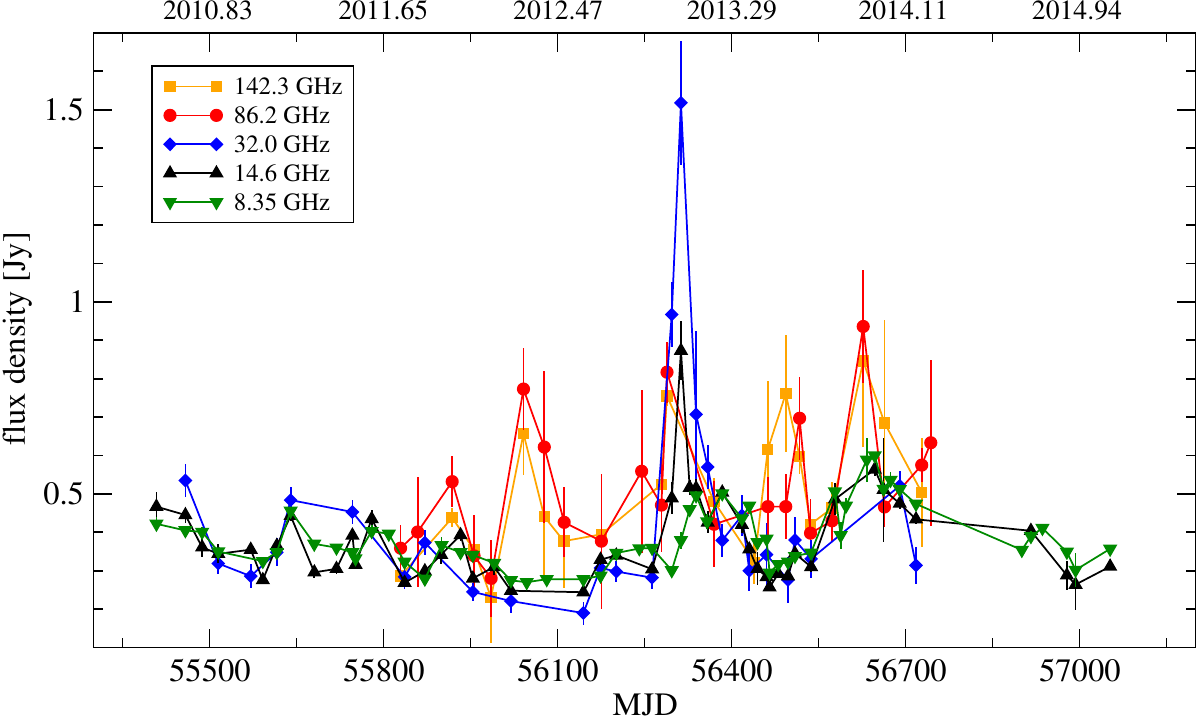}\\
\vspace{0.2cm}
\includegraphics[width=0.6\linewidth]{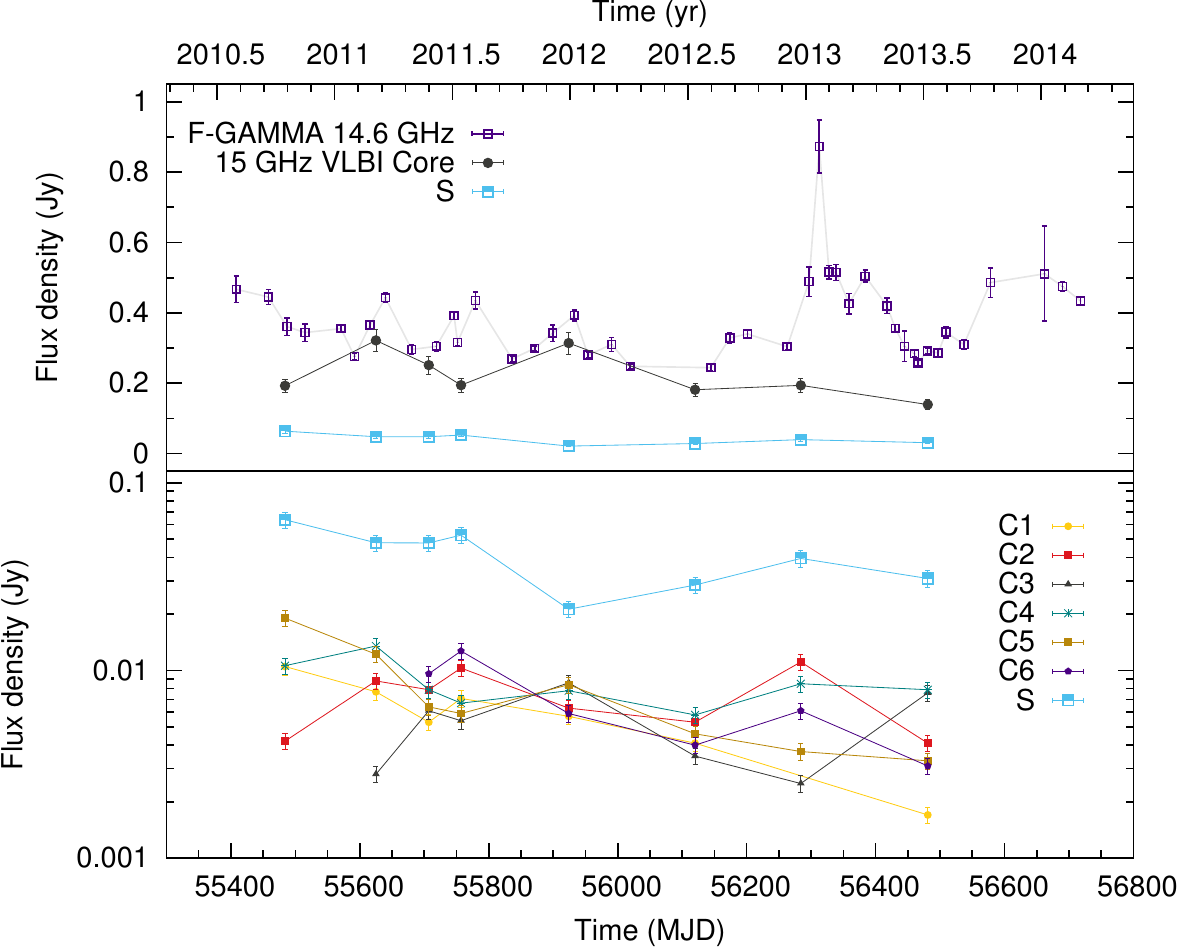}

\caption{Single dish and VLBI radio light curves of 1H\,0323+342. 
The top panel features the F-GAMMA
total flux density light curves at 142.3, 86.2, 32.0, 14.6 and
8.35\,GHz. In the middle panel, those at 14.6\,GHz from EB (purple squares)
along with that of the 15\,GHz VLBI core (black) and the stationary component S
(light blue) are shown. The bottom panel features the light curves of all
individual VLBI components. Note the different scales and the presence of 
component S in both panels for visual comparison.}
\label{fgamma_lcs}
\end{figure}

Example light curves of 1H\,0323+342 at the selected frequencies of 142.3, 
86.2,
32.0, 14.6 and 8.35\,GHz, covering the period 2010 July (2010.6;
MJD\,55408) to 2015 January (2015.1; MJD\,57053), are shown in
Fig. \ref{fgamma_lcs} (top panel).
Earlier results of this NLS1 monitoring campaign have been presented in e.g.
\cite{2011nlsg.confE..26F} and \cite{2012A&A...548A.106F}. The detailed
multi-frequency analysis and results of the full NLS1 radio data sets (five
sources) obtained between 2010 and 2014 are presented and discussed in
\cite{2015A&A...575A..55A}.

\subsection{VLBI monitoring}\label{Sec:Data:Mojave}

The MOJAVE program \citep{2009AJ....137.3718L} has been monitoring
1H\,0323+342 with the Very Long baseline Array (VLBA) at 15\,GHz between 2010
October and 2013 July, and a total of eight observing epochs of fully
self-calibrated data sets are
 available\footnote{\url{www.physics.purdue.edu/MOJAVE/}}.

The brightness distribution of the source at each epoch has been
parameterised by sets of two-dimensional circular Gaussian functions
with the \textsc{DIFMAP} package \citep[][]{1994BAAS...26..987S}. This uses
a $\chi^{2}$ minimisation  algorithm (\textsc{MODELFIT}) to fit the parameters
of the Gaussian components -- i.e. flux density, radial separation from the
phase center, position angle (PA) and size -- to the visibility function.

First, the image of each epoch was analyzed independently by building the model
iteratively, i.e. adding new components as long as the residual image
shows significant flux density and the new component improved the
overall fit in terms of the $\chi^{2}$ value. The addition of a
component was followed by iterations of \textsc{MODELFIT} (until
the fit has converged) and self-calibration of the visibility
phases.
In addition, a cross-check of our model was performed by
comparing the position of the components with the brightness distribution 
in the \textsc{CLEAN} image. We adopt an uncertainty of 10\% for the flux
density of components \citep[e.g.][]{2009AJ....137.3718L}. The
positional uncertainty is set to 1/5 of the beam, if they are
unresolved, or 1/5 of their FWHM for those resolved. 
A few data points deviate from the fitted lines, as common of such type of 
studies, 
attributable to possible epoch-to-epoch illumination of different emission 
regions within 
the same knot, and small differences in the uv-coverage.
The uncertainty in PA
is given by $\Delta {\rm PA} = \arctan(\Delta {\rm X}/r)$, with $\Delta {\rm 
X}$
the positional uncertainty and $r$ their separation from the core
\cite[e.g.][]{2016A&A...586A..60K}.
We note that neither during the cleaning nor the
model-fitting process, significant flux density could be associated with a
putative counter-jet above the noise-level of a given epoch (see 
Sect. \ref{counter_jet_ratio}).
Details of the resulting multi-epoch model parameters are given in 
the Appendix (Table \ref{tab:modelfit}). All individual \textsc{MODELFIT} maps
 are shown in Fig. \ref{vlbi_images}.

\section{Results from total flux density monitoring}\label{D_var}

The single-dish light curves\footnote{Single-dish multi-frequency data from
the F-GAMMA program covering the period MJD\,55400--56720
\citep{2015A&A...575A..55A} are publicly available via
\url{http://cdsarc.u-strasbg.fr/viz-bin/qcat?J/A+A/575/A55}} displayed in Fig.
\ref{fgamma_lcs} (top) show 1H\,0323+342 in a phase of mild,
low-amplitude but rather fast
variability during the period $\sim$2010 to 2012. Here, the
14.6\,GHz light curve, for instance, exhibits repeated 
low-amplitude flares with amplitudes ($\Delta S =
S_{\rm{max}}-S_{\rm{min}}$) of $\sim$120--170\,mJy and time
scales ($\Delta t=t_{\rm{max}}-t_{\rm{min}}$) of
$\lesssim$40--100\,days. Using these values, we calculate lower
limits for the apparent brightness temperature $T_{\rm{B,var}}$
associated with these events ranging
between $5\times10^{10}$ and $4\times10^{11}$\,K.

Starting in 2012, the source showed more prominent activity with a
first larger mm-band flare peaking around 2012.2 and subsequently, a
large outburst occurring between $\sim$2012.6 and 2013.5. The
latter appears to exhibit 2--3 sub-flares with the most prominent
event showing a variability amplitude of $\Delta S\sim\,570$\,mJy
during a time period of about 50\,days at 14.6\,GHz. This yields a
lower limit for $T_{\rm{B,var}}$ of about $9\times10^{11}$\,K. The observed
brightness temperature can be further constrained using the highest 
significant amplitude variation occuring at adjacent times, 
between MJD 56296.8 and 
56312.7 in the 14.6\,GHz light curve (Fig.
\ref{fgamma_lcs}). This corresponds to a variability amplitude of 384\,mJy and
a time scale of 15.9 days. These figures yield  an 
apparent
$T_{\rm{B,var}}$ of $5.7\times10^{12}$\,K.

Assuming an intrinsic equipartition brightness temperature limit of
$5\times10^{10}$\,K \citep{Readhead1994ApJ}, we can estimate lower limits for
the variability Doppler factor $D_{\rm{var}}$ required to explain
the excess of $T_{\rm{B,var}}$ over the theoretical limit
\citep[e.g.][]{2008A&A...490.1019F, 2009A&A...494..527H, 
2015A&A...575A..55A}. 
The maximum observed value of $5.7\times10^{12}$\,K then
yields $D_{\rm{var}}$\,$\gtrsim$\,5.2 for the 14.6\,GHz variability
seen in Fig. \ref{fgamma_lcs}.
Based on \textit{a different method}, their flare decomposition 
method, \cite{2015A&A...575A..55A} deduce a $D_{\rm{var}}$\,$\gtrsim$\,3.6 
for the flare seen at 14.6\,GHz, peaking at MJD $\sim$56300.
For a full event-by-event treatment of the data sets until Spring 2014 see the 
aforementioned publication.

\section{Results from VLBI monitoring}\label{kinematics}

\begin{figure}[h!]
    \centering
    \includegraphics[width=0.55\linewidth]{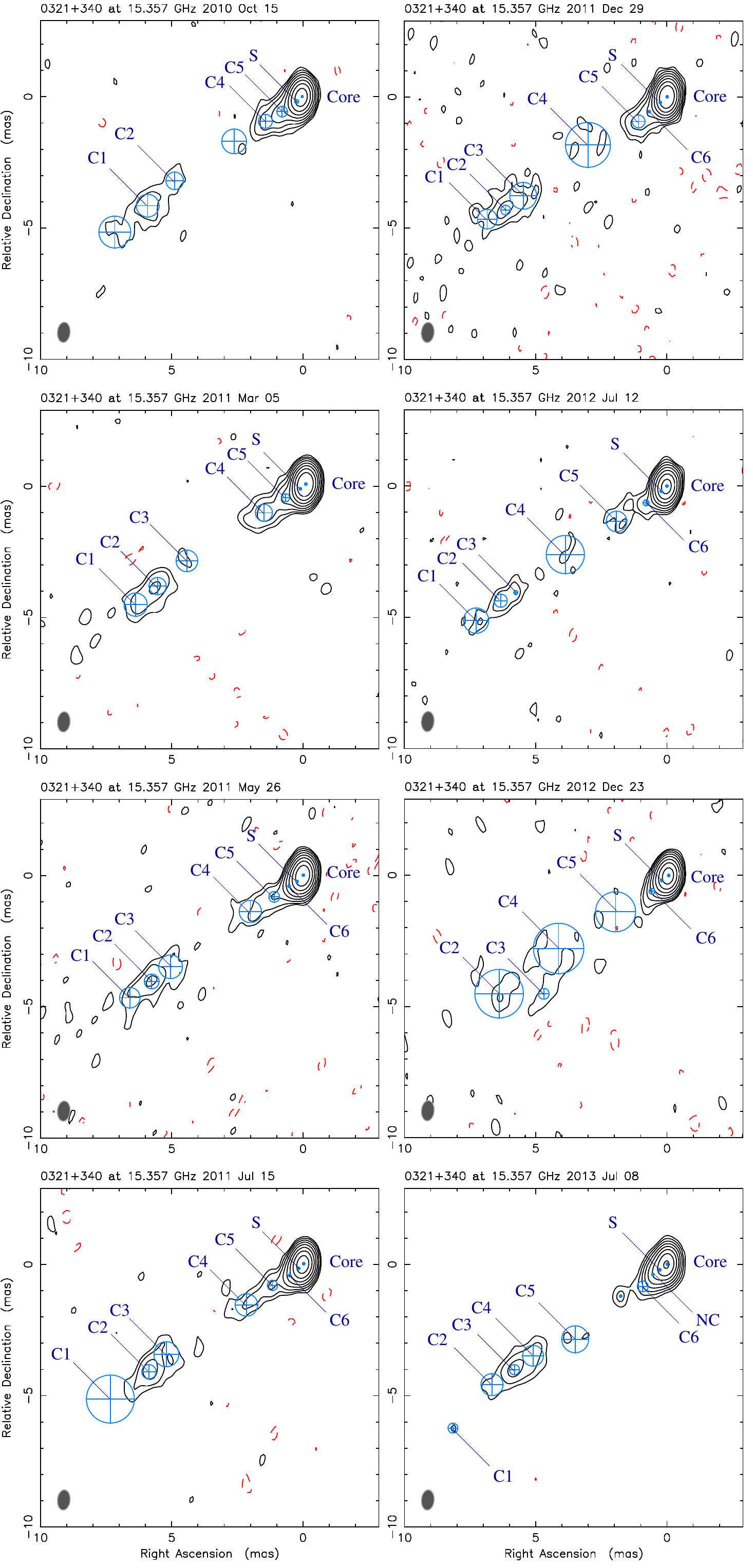}
    \caption{Eight uniformly-weighted \textsc{MODELFIT} maps of 1H\,0323+342
from the MOJAVE observations at 15\,GHz. Contour levels correspond to $-0.3$\%,
0.3\%, 0.6\%, 1.2\%, 2.4\%, 4.8\%, 9.6\%, 19.2\% and 38.4\% of the highest peak
flux density of 0.344 Jy\,beam$^{-1}$ (epoch 2011.17) for common reference. All
maps are convolved with an average beam with major and minor axes of 0.77 and
0.48\,mas, respectively, with the major axis at PA $= -5^{\circ}$.}
          \label{vlbi_images}
 \end{figure}

Figure \ref{vlbi_images} shows the final \textsc{MODELFIT} maps of
1H\,0323+342 at $15$\,GHz. On parsec scales, the
source is characterized by a one-sided jet that appears remarkably straight,
oriented at a mean position angle of ${\sim} 124^{\circ}$, and extending
out to almost $10$\,mas from its bright, unresolved core. The latter
contains more than 70\% of the jet's total flux density (see
Fig. \ref{fgamma_lcs}).

As also seen in the VLBI maps of Fig. \ref{vlbi_images}, the jet of
1H\,0323+342 is not undisrupted all the way until its maximum
angular extent. Further downstream from the core
the jet exhibits an area  of reduced surface brightness (a gap of emission).
On average this 
is visible between $3$ to $6$\,mas away from the core at the
PA of the jet, whereafter the flow becomes visible again. This effect could
be attributed to the weighting of the VLBI data, selected such that resolution
is maximized at the expense of sensitivity to putative extended
structure\footnote{For a stacked image of all naturally-weighted CLEAN maps
from the MOJAVE survey see www.physics.purdue.edu/astro/MOJAVE/stackedimages/
0321+340.u.stacked.icn.gif}.

At our observing frequency the jet can be decomposed into six to nine
\textsc{MODELFIT} components.
Based on the parameters of the model and their temporal evolution we are able 
to
positively cross-identify seven components between all eight observing
epochs. Starting from the outermost component -- in terms of radial
separation from the core -- we use the letter C followed by a number
between $1$ and $6$ to designate and hereafter refer to them. Apart
from travelling components, our findings suggest the presence of a
quasi-stationary component, referred to with the letter S, positioned
very close to the core at a mean distance of ${\sim}
0.3$\,mas. 
Furthermore, at epoch $2013.52$ a new component appears to emerge, most
probably ejected at some time between the latest two observing epochs; i.e.
between $2012.98$ and $2013.52$ (see Sect. \ref{VLBI_discussion}). We refer to
this knot as NC which stands for `new component' (cf. VLBI maps in Fig.
\ref{vlbi_images}).

\begin{figure}
    \centering
   \includegraphics[width=0.6\linewidth]{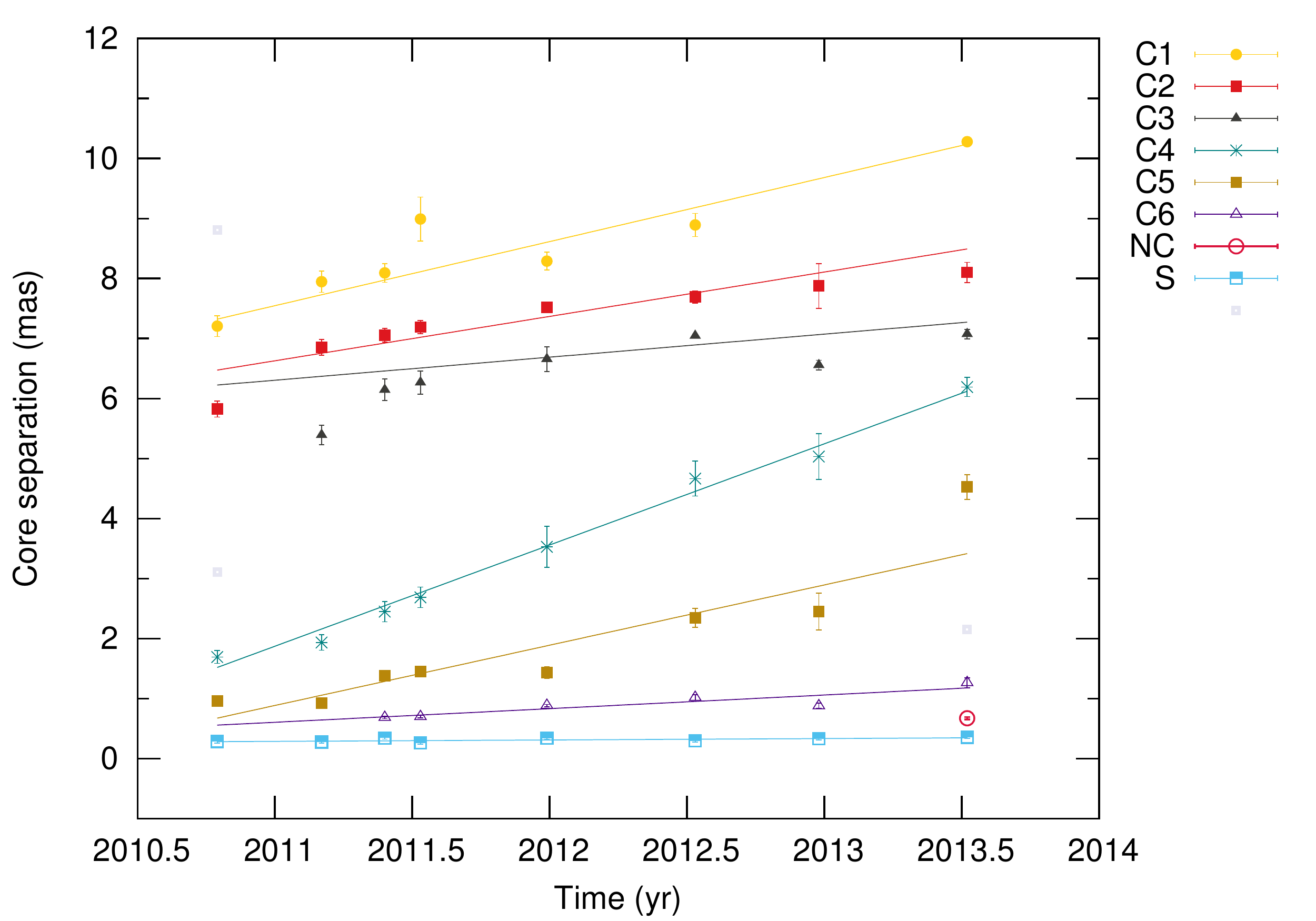}
    \caption{Temporal evolution of fitted component separation from the core.  
  Line segments represent the best-fit linear model describing each component's
  kinematical behavior. Apart from robustly-identified knots (C1--C6), open
  squares denote non-robust features. The new component (NC) visible in the
  latest epoch only, is shown as a red circle immediately after
  the quasi-stationary feature S.}
    \label{fig:CoreSep}
 \end{figure}

The kinematical properties of the jet were derived by identifying model
components throughout our multi-epoch observations, which describe the 
same jet features. 
The kinematical parameters for each knot are obtained through a weighted linear
regression fit of their positions relative to the assumed stationary core over
time. In Fig. \ref{fig:CoreSep} each component's radial separation from the 
core
at each observing epoch is shown, along with the linear fits. From the slope of
each fit we deduce the component's proper motion $\mu$ relative to the core and
its apparent velocity $\beta_{\rm app}$ in units of the speed of light. The
results are summarized in Table \ref{tab:kinematics}.  

Beyond the first mas and before the area of reduced radio emission,
$\beta_{\rm app}$ shows an increasing trend. There, superluminal knots C5 and 
C4
travel with apparent velocities of ${\sim} 4$\,c and ${\sim} 7$\,c,
respectively. Superluminal knot C4 is the fastest moving component in
the relativistic flow of 1H\,0323+342.

After the gap of emission, when the jet becomes visible again, the flow
is characterized by slower speeds. Components C3, C2 and C1 are
used to describe the elongated area of emission extending at a radial
distance between 6 and 10\,mas away from the core. Their speeds are
superluminal, only slightly slower than in the jet area before the
gap, with $\beta_{\rm app,\,C3} \simeq 1.6$, $\beta_{\rm app,\,C2}
\simeq 3$, and $\beta_{\rm app,\,C1} \simeq 4$.

Ejection times of knots are derived by
back-extrapolating their motion to the time of zero separation from
the core and are given in the last column of Table
\ref{tab:kinematics}. The source appears very active in this respect,
showing an ejection of a new VLBI component every ${\sim}3.1$ years on
average with the earliest ejection in $\sim$1994.6 and the most recent
component ejections in 2009/2010.

\begin{table}
     \centering
     \caption[]{Kinematical parameters for all components of
1H\,0323+342. Columns from left to right are: (1) component designation, (2)
proper motion in mas/yr, (3) apparent velocity in units of c, (4) time of knot
ejection.}
     \label{tab:kinematics}
     \begin{tabular}{cccccccccccc}
     \hline
 ID    &     $\mu$         &  $\beta_{\rm app}$  &        $t_{\rm ej}$     \\
     &   $({\rm mas\,yr}^{-1})$ &    $({\rm c})$  &   $({\rm yr})$\\
      \hline
         C1     & $1.07 \pm 0.09$ & $4.38 \pm 0.35$ & $2003.9 \pm  0.7$ \\    
         C2     & $0.74 \pm 0.15$ & $3.03 \pm 0.63$ & $2002.0 \pm  2.1$ \\    
         C3     & $0.38 \pm 0.26$ & $1.58 \pm 1.06$ & $1994.6 \pm 12.1$ \\    
         C4     & $1.69 \pm 0.07$ & $6.92 \pm 0.29$ & $2009.9 \pm  0.1$ \\    
         C5     & $1.00 \pm 0.20$ & $4.12 \pm 0.82$ & $2010.1 \pm  0.3$ \\    
         C6     & $0.23 \pm 0.06$ & $0.93 \pm 0.24$ & $2008.3 \pm  0.9$ \\    
          S     & $0.02 \pm 0.01$ & $0.10 \pm 0.05$ &         \dots     \\  
         \hline        
     \end{tabular}
 \end{table}


The main variable component at $15$\,GHz, according to our VLBI
findings, appears to be the core whereas all other VLBI jet components
of 1H\,0323+342 share a lower 
flux density level. Comparison between the F-GAMMA filled-aperture light curve
at $14.6$\,GHz and the VLBI total flux density of all components (see
Fig. \ref{fgamma_lcs}) indicates that VLBI flux densities account for
most of the single-dish flux density, a hint that extended structure
does not contribute a substancial flux density level. It in fact suggests
that the total intensity flares originate in the unresolved $15$\,GHz
VLBI core region of the source, since the core flux density is variable and
accounts for more than 70\% of the observed total VLBI flux density.

\section{Jet-to-counter-jet ratio}\label{counter_jet_ratio}

Assuming intrinsic symmetry between jet and counter-jet, 
the jet-to-counter-jet ratio $R$ represents the ratio of flux
densities of the approaching and receding jets and is given by
\begin{equation}
    R = \dfrac{S_{\rm Jet}}{S_{\rm cJet}} = 
    \left( \dfrac{1 + \beta \cos \theta}{1 - \beta \cos \theta} \right)^{2 -
\alpha} = \left(\beta_{\rm app}^{2} + D^{2} \right)^{2-\alpha},
\label{eq_Ratio}
\end{equation}
where $\beta$ and $\theta$ are the intrinsic speed and jet angle to
the observer's line of sight. The index ($2-\alpha$) refers to a
continuous jet flow and changes to ($3-\alpha$) for a `blobby' jet
\citep{1979Natur.277..182S}.

For the NLS1 galaxy 1H\,0323+342 the total flux density
of all components in the VLBA images is associated with the jet, while
putative counter-jet emission from the source is not visible at the opposite
direction of the core and to the North (see Fig. \ref{vlbi_images};
this statement assumes, that there is no contribution from counter-jet emission
to the brightest radio component). Despite the
fact that there is no sign of emission from the counter-jet side, we have
nonetheless tried to constrain its non-observability. First,
the VLBI map with the best signal-to-noise ratio was selected, i.e. epoch
2011 March 5. Using the \textsc{CLEAN} algorithm \citep{1974A&AS...15..417H}
and by utilizing \textsc{CLEAN} windows to constrain the area where the
algorithm searches for peaks, we have tried to place delta components
at the opposite side of the jet in an attempts to `artificially
create' a fake counter-jet. Mostly positive and negative components
with no significant flux (below the noise level of the map) were
fitted in this area of the image -- if at all -- as expected. Another
test was performed using the \textsc{MODELFIT} algorithm, this time by
attempting to fit the self-calibrated (only for the phase term) visibilities
with circular Gaussian components. The results were negative in this case
too, with no significant components on the opposite side of the
jet. Under the assumption that Doppler de-boosting is responsible for
the suppression of the counter-jet one can obtain a lower limit for
the jet-to-counter-jet ratio $R$.
We calculate R using the \textsc{MODELFIT} map at epoch 2011 March 5.
As an upper limit for the flux density of the counter-jet, $S_{\rm cJet}$, we 
take three times the rms integrated flux density, over a region with comparable 
size to and in the opposite side of the jet (rms $= 0.38$\,mJy). Being 
undetected, any counter-jet emission ought to be below this level.
Consequently, through $R = S_{\rm Jet}/S_{\rm cJet} = S_{\rm
Jet}/S_{3 \sigma}$ , we estimate that $R \ge 363.5$, 
where $S_{\rm Jet}$ is the 
total integrated flux density of the fitted core and jet components.

A first estimate for the viewing angle towards the source can then be
obtained through
\begin{equation}
    \theta \le \arccos \left( \frac{R^{\frac{1}{2-\alpha}} -
1}{R^{\frac{1}{2-\alpha}} + 1} \right) \hspace{0.2cm} \beta^{-1}.
\end{equation}
Under the assumption that the intrinsic speed of the components of the pc-scale
jet is close to the speed of light, i.e. letting $\beta \rightarrow 1$, and 
with a mean $\alpha = -0.04$ \citep{2015A&A...575A..55A}, we arrive at an upper 
limit for the viewing
angle of $\theta \le 27^{\circ}$, for $R = 363.5$.  If instead, we exclude
the core flux density for this calculation, an even weaker constraint is
obtained. Note that solving Eq. \ref{eq_Ratio} for $D$, 
this value of $R$, combined with the highest measured apparent velocity in the 
jet, implies a non-physical Doppler factor ($D \in \mathbb{I}$; imaginary).
To remedy this and get a real-valued Doppler
factor, a much higher jet-to-counter-jet ratio is needed (${\sim} 2700$), which
in turn yields a viewing angle of $\theta \le 16^{\circ}$.

Another way of constraining the viewing angle towards the source is
the following. Having a measured value for $\beta_{\rm app}$ we
calculate the minimum Lorentz factor in order to observe this apparent
speed as
\begin{equation}
    \gamma_{\rm min} = \sqrt{\beta_{\rm app}^{2} + 1}.
    \label{gamma_min}
\end{equation}
Then using Eq. \ref{gamma_min} under the assumption that this is the
Lorentz factor characterizing the flow, the critical viewing angle is
estimated through \citep[][]{1966Natur.211..468R}
\begin{equation}
    \theta_{\rm crit} = \arcsin \left( \dfrac{1}{\gamma_{\rm min}} \right).
\end{equation}
In this way and for $\beta_{\rm app} = 6.92$, the minimum Lorentz factor is
$\gamma_{\rm min} = 6.99$ and correspondingly we obtain 
$\theta_{\rm crit} = 8.2^{\circ}$. For a discussion on the use of the
critical angle as a proxy for the viewing angle see \cite{2016A&A...586A..60K}.

\section{The viewing angle towards 1H\,0323+342}\label{sect:Combi}

A viewing angle estimate towards 1H\,0323+342 can be obtained based on our
radio variability and imaging results. With regard to the obtained  
Doppler factor, we would first like to note, that there is, in principle, 
several methods for estimating Doppler factors in blazars, and NLS1 galaxies in 
particular.
One approach is based on SED modelling, which has also been carried out for
1H\,0323+342 (e.g.,
\citep{
2009ApJ...707L.142A,
2013ApJ...774L...5Z,
2014ApJ...789..143P,
2015AJ....150...23Y,
2015ApJ...798...43S}
While this is a very important approach, the estimated parameters
still come with a number of uncertainties, based on, for instance, 
non-simultaneity of some of the data points, assumptions about black holes mass 
and  correspondingly the accretion disk contribution to the SED, and 
based on different assumptions of different authors on the dominant 
contribution to EC processes (like the BLR or the molecular torus, or both).
Alternatively, the variability Doppler factor in the radio regime
can be directly estimated from radio flux density monitoring data. We have 
carried out such a program over several years, and it is these data sets we 
have used here for directly  estimating the variability Doppler factor (the 
value we find (Sect. 3) is consistent within the errors with estimates based on 
SED fitting \citep[e.g.,][their Tab. 6]{2015AJ....150...23Y}. We have 
performed the estimate at 15 GHz, because this is the frequency at which the 
VLBI data were taken.

Combining the variability Doppler factor $D_{\rm var}$ estimated in Sect.
\ref{D_var}, based on our single-dish monitoring with the apparent jet
speed $\beta_{\rm app}$ estimated in Sect. \ref{kinematics}, we are able to
estimate the variability Lorentz factor $\gamma_{\rm var}$ and the viewing
angle $\theta_{\rm var}$ towards 1H\,0323+342 according to
\citep[][for the derivation see
\citealp{2013Ap&SS.348..193O}]{2005AJ....130.1418J}
\begin{equation}
\gamma_{\rm var} = \dfrac{ \beta_{\rm app}^{2} + D_{\rm var}^{2} + 1 }{ 2
D_{\rm var} } ~~~ \textrm{and}
\label{eq:minLorfctr}
\end{equation}

\begin{equation}
\theta_{\rm var} = \arctan \left( \dfrac{ 2 \beta_{\rm app} }{ \beta_{\rm
app}^{2} + D_{\rm var}^{2} - 1 } \right).
\label{eq:thetaVar}
\end{equation}

Ideally, one aims at using the apparent jet speed of the travelling
VLBI knot causing the outburst for which $D_{\rm var}$
has been obtained. However, this is not possible in the case of
1H\,0323+342, given the available data sets (see also Sect.
\ref{VLBI_discussion}). Nevertheless, we can assume the range of obtained
$D_{\rm var}$ and jet speeds in Sect. \ref{D_var} and \ref{kinematics} as
representative, average jet characteristics to obtain a range of plausible
values for the Lorentz factor as well as viewing angle of the source.

Assuming a constant Doppler factor, we use Eqs. \ref{eq:minLorfctr} and
\ref{eq:thetaVar} and combine the observed
apparent velocities of the flow with the variability Doppler factors discussed
in Sect. \ref{D_var}. While the most stringent constraint is obtained from the
combination of the lowest observed $\beta_{\rm app} = 0.93$ with the highest
Doppler factor, given the large scatter of observed speeds, we perform the same
calculation using the lowest $D_{\rm var} \ge 3.6$ and the highest $\beta_{\rm
app} = 6.92$. In the former case we obtain a variability Lorentz factor
$\gamma_{\rm var} \ge 2.8$ and the viewing angle $\theta_{\rm var} \le
4.0^{\circ}$. In the latter, we have 
$\gamma_{\rm var} \ge 8.6$ and $\theta_{\rm var} \le 13.0^{\circ}$. The final
range of values for the viewing angle is $\theta_{\rm var} \le
4^{\circ}$--$13^{\circ}$.
Results of all calculations are summarized in Table \ref{tab:resultsAll}.
\begin{table}
\centering
\caption{
  Physical parameters of the jet of 1H\,0323+342. The left column contains the
measured quantities, used in each calculation, leading to the values in the
right column. 
  The table from top to bottom contains the results with the use of VLBI data
only, the critical values for the Lorentz factor (minimum value) and the 
viewing
angle, and finally, results obtained through the combination of VLBI and
single-dish data.}
\begin{tabular}{l r}
\hline
 Measured quantities       &   Inferred parameters \\ 
\hline
\noalign{\smallskip}
\multicolumn{2}{c}{VLBI only}                       \\
\noalign{\smallskip}
 $R = 363.5$               & $D \in \mathbb{I}$     \\
 $\beta \rightarrow 1$     & $\theta \leq 27^{\circ}$   \\  
\noalign{\medskip}
 $R_{\rm req}$ = $2700$    & $D \in \mathbb{R}$      \\
 $\beta \rightarrow 1$     & $\theta \leq 16^{\circ}$	  \\
\hline
\noalign{\smallskip}
\multicolumn{2}{c}{Critical values}                     \\
\noalign{\smallskip}
 $\beta_{\rm app} = 6.92$   & $\gamma_{\rm min} = 6.99$ \\
                           & $\theta_{\rm crit} = 8.2^{\circ}$ \\ 
\hline
\noalign{\smallskip}
\multicolumn{2}{c}{VLBI and single dish}                \\
\noalign{\smallskip}
 $\beta_{\rm app, min} = 0.93$  & $\gamma_{\rm var} \ge 2.8$ \\
 $D_{\rm var} \ge 5.2$     & $\theta_{\rm var} \le 4^{\circ}$ \\
\noalign{\medskip}
 $\beta_{\rm app, max} = 6.92$   & $\gamma_{\rm var} \ge 8.6 $         \\
 $D_{\rm var} \ge 3.6$     & $\theta_{\rm var} \le 13^{\circ}$ \\
\hline
\end{tabular}
\label{tab:resultsAll}
\end{table}

\section{Discussion}\label{sect:Disc}

\subsection{On the variability of 1H\,0323+342}\label{VLBI_discussion}

The unresolved 15\,GHz VLBI core flux density is variable and accounts 
for more than 70\,\% of the total VLBI flux density. Consequently, 
it is likely the region responsible for the variability 
and flares seen in the single-dish total flux density light curves,
however, a direct connection between total flux density outbursts 
and pc-scale VLBI jet features cannot be established given the poor VLBI 
data sampling. This is particularly true for the most dramatic 
outburst observed around 2013.1. In addition, given the limited time 
baseline of the single-dish observations, we are unable to establish 
a one-to-one correspondence between component ejection times and total 
intensity flares. This becomes clear from the estimated ejection times 
of components reported in Table \ref{tab:kinematics}, none of which 
appears to have separated from the core within the time baseline of
our single-dish radio monitoring. The only exception appears to be a new
jet component (refereed to as NC) only seen to be present in the data of the
last epoch ($2013.52$, see Fig. \ref{fig:CoreSep}) with an integrated 
flux density of ${\sim} 19$\,mJy. Given the flaring activity
preceding its separation from the core, this knot could be tentatively
associated with the above mentioned, most dramatic outburst seen in our 
single-dish data around epoch ${\sim} 2013.1$ (see Fig.~\ref{fgamma_lcs}). 

\subsection{On the viewing angle towards 1H\,0323+342}

A first indication that the viewing angle towards 1H\,0323+342
is not exceptionally small is the remarkably straight jet. All jet
components lie at a mean position angle of about 124 degrees, with minor
deviations. Since the jet appears so aligned, this likely implies that the
viewing angle is not too small, otherwise PA variations would be greatly
enhanced and more pronounced due to relativistic aberration effects, as in the
case of objects seen at very small viewing angles (${\sim}
1^{\circ}$--$2^{\circ}$).

As shown in Sect. \ref{counter_jet_ratio}, the viewing angle $\theta$ towards
1H\,0323+342  can be loosely constrained by using the interferometric data
alone. Here, the  minimum Lorentz factor estimate yields a value of
$8.2^{\circ}$ from $\theta_{\rm crit}$, while the jet-to-counter-jet ratio
estimate provides loose upper limits of $\theta\,\le\,27^{\circ}$ and 
ultimately
$\theta\,\le\,16^{\circ}$.

The combination of VLBI kinematics and single-dish variability allows us to 
further
constrain $\theta$ (Sect. \ref{sect:Combi}). We note that the variability
Doppler factors constitute only  lower limits, given the fact that the emission
region can be smaller than the size implied by the variability time scales
obtained.
\cite{2015A&A...575A..55A} presented a detailed
analysis of the single-dish data sets of 1H\,0323+342 between 2010.6 and 2014.2
based on a multi-frequency flare decomposition method. At 14.6\,GHz, their 
study
yielded $D_{{\rm var}} \ge 3.6$. Consequently, using
$D_{{\rm var}}>3.6$ and $\beta_{\rm app} = 6.92$\,c we obtain a viewing
angle of ${\le} 13^{\circ}$.

Additionally, in Sect. \ref{D_var} we deduced a variability Doppler factor
$D_{\rm var} \ge 5.2$. This, combined with the slowest apparent speed of the
moving knot C6, yields another, more stringent upper limit for $\theta$ of
${\le}4^{\circ}$. The very small apparent speed of C6 can be reconciled with
such a high Doppler factor, if projection effects are at play and the jet is
pointed close to the observer's line of sight, as seen from the inferred
aspect angle.

\subsection{The viewing angle towards the NLS1 galaxy SBS\,0846+513}\label{SBS}


Using the results presented by \cite{2013MNRAS.436..191D} \citep[see
also][]{2012MNRAS.426..317D} we can also estimate the viewing angle
towards the radio and $\gamma$-ray-loud NLS1 SBS\,0846+513, using the same
method presented in Sect. \ref{sect:Combi}. The authors report an apparent
jet speed for this source of 9.3\,c. Furthermore, we estimate $D_{\rm var}$ as
in Sect. \ref{D_var} from the 15\,GHz variability presented in
Figs. 3 and 4 of \cite{2012MNRAS.426..317D} to be of the order ${\ge} 6$--$7$,
in agreement with a value of ${\ge} 7.1$ which we obtain from the variability
characteristics given in \cite{2013MNRAS.436..191D} and using an
intrinsic equipartition brightness temperature limit of
$5\times10^{10}$\,K. These values for $\beta_{{\rm app}}$ and $D_{\rm var}$
then provide a viewing angle estimate of ${\le}8^{\circ}$--$9^{\circ}$ using
Eq. \ref{eq:thetaVar}.

\subsection{Comparison with classical blazars}

Out of the six moving features populating the jet of 1H\,0323+342, five are
superluminal and only one moves with a marginally subluminal velocity. The
apparent velocities span a wide range between 0.93\,c up to 6.92\,c. Such a
range of apparent speeds is often observed for the class of
blazars \citep[e.g.][]{2013AJ....146..120L}.

The Lorentz factors obtained for 1H\,0323+342 and SBS\,0846+513 are in the
observed range of blazars
but at the lower end of the blazar Lorentz factor distribution and more
typical for BL Lac objects
\citep[e.g.][]{2009A&A...494..527H, 2010A&A...512A..24S}.

The different methods we employ constrain the viewing angle
towards 1H\,0323+342 to $\theta \leq 4^{\circ}$--$13^{\circ}$,
while $\theta \leq
8^{\circ}$--$9^{\circ}$ for SBS\,0846+513. These numbers are larger than the
average viewing angles towards blazars, of ${\sim} 1^{\circ}$--$2^{\circ}$
\citep[e.g.][]{2009A&A...494..527H, 2010A&A...512A..24S}, but given that out
measurements are upper limits, the angles could still be consistent with
smaller values. Whether this is a general characteristic of radio-loud,
$\gamma$-ray-loud NLS1 galaxies needs to be answered by detailed studies of a
larger sample of this source type.

Based on VLA observations of the kpc-scale jets of three radio-loud but
non-$\gamma$-ray detected NLS1 galaxies, the finding of
\cite{2015ApJ...800L...8R} suggest that these sources are seen at a
moderate angle to the line of sight (10$^{\circ}$--15$^{\circ}$).

Our results confirm previous findings that radio-loud, $\gamma$-ray-loud NLS1
galaxies, as a class, share blazar-like properties\footnote{Note a few
possible exceptions, in the following sense: while the $\gamma$-bright source
PKS\,2004-447 shows many blazar-like characteristics, the angle of the jet to
the line of sight is likely fairly large as indicated by the persistent steep
radio spectrum and the diffuse emission on the counter-jet side
\citep[][]{2015arXiv150903735K, 2015arXiv151102631S}. Further exceptions may be
the candidate $\gamma$ NLS1 galaxies B3\,1441+476  and RX\,J2314.9+2243, which 
both show
steep radio spectra \citep[][respectively]{2015arXiv151005584L,
2015A&A...574A.121K}.}.
However, compared to typical GeV-blazars
\citep[e.g.,][]{2012ApJ...752..157Z,2014ApJ...788..104Z},
and references therein), 
most radio-loud NLS1 galaxies appear
as lower brightness temperature sources\footnote{See
\cite{2013MNRAS.436..191D} for an exception.}, possibly
seen at slightly larger viewing angles than classical blazars. These properties
can then explain the typically small Doppler boosting factors and low flux
densities of these sources \citep[e.g.][]{2010arXiv1007.0348F,
2015A&A...575A..55A, 2015A&A...575A..13F, 2015ApJS..221....3G}. The application
of the method presented here to a larger sample of radio-loud NLS1 galaxies 
will
provide us with first robust measurements of the viewing angle towards this
population of AGN.

\section{Summary and conclusions}\label{sect:summary}

We have presented a VLBI study of the $\gamma$-ray- and radio-loud NLS1 galaxy
1H\,0323+342, including a viewing angle measurement towards the
inner jet of this source, by combining single-dish and VLBI monitoring. The
salient results of our study can be summarized as follows:

\begin{enumerate}
  \item 1H\,0323+342 displays a pc-scale morphology of the core-jet type. The
emission is dominated by the unresolved core which contains more than 70\% of
the jet's total flux density. The jet extends straight up to about 10\,mas from
the bright nucleus, confirming previous results at a lower frequency. The
VLBI structure at 15\,GHz can be represented using six moving components and 
one
quasi-stationary feature.
 
 \item Five out of the six moving knots exhibit superluminal motion. The range
of apparent velocities is 0.93\,c up to 6.92\,c. 
Closer to the core (${\le} 1$\,mas) speeds are
subluminal. Between 1\,mas and 5\,mas, the highest speeds of the flow are
observed with components moving at about 4\,c and 7\,c.

 \item 1H\,0323+342 is active, showing flux density flares, of low to mild
amplitude, but very rapid compared to the typical long-term variability of
blazars. The behaviour is seen both in single-dish total flux
density and on VLBI scales. Previous findings infer high brightness
temperatures (above the equipartition limit) and a variability Doppler factor 
of
3.6, taking into account the parameters of full flaring episodes. Here we also
deduce an extreme $T_{\rm B,\,var} = 5.7\times10^{12}$\,K and $D_{\rm var} =
5.2$, based on the most rapid variability of the highest amplitude.

 \item Using different methods we ultimately constrain the viewing angle
towards 1H\,0323+342 to be in the range $\theta_{\rm var}
\le 4^{\circ}$--$13^{\circ}$.

 \item Applying similar estimates to published results of another radio-loud
NLS1 galaxy, SBS\,0846+513, we estimate a viewing angle of ${\le}
8^{\circ}$--$9^{\circ}$.
\end{enumerate}

Both sources are therefore seen at a small viewing angle.
Taken at face value, inferred numbers are larger than the bulk of the blazar
population, but still consistent with each other, given that our estimates are
upper limits. Aspect angle measurements of a larger sample
of radio-loud NLS1 galaxies will provide us with important new insight into
orientation scenarios for NLS1 galaxies.

\normalem
\begin{acknowledgements}
We  would like to thank our referee for very useful comments and suggestions.
The authors thank H. Ungerechts and A. Sievers for their support
in the course of the IRAM 30-m observations, and B. Boccardi for carefully 
reading the manuscript.
V.K., I.N. and I.M. were supported for this research through a stipend from the
International Max Planck Research School (IMPRS) for Astronomy and Astrophysics
at the Universities of Bonn and Cologne. R.S. was supported by the Deutsche
Forschungsgemeinschaft grant WI 1860/10-1.
E.R. acknowledges partial support by the the Spanish MINECO project
AYA2012-38491-C02-01and by the Generalitat Valenciana project
PROMETEOII/2014/057.
This research made use of observations with the 100-m telescope of the
MPIfR (Max-Planck-Institut f\"ur Radioastronomie) at Effelsberg, and
observations with the IRAM 30-m telescope. IRAM is supported by INSU/CNRS
(France), MPG (Germany) and IGN (Spain).
This research has made use of data from the MOJAVE database that is
maintained by the MOJAVE team (Lister et al., 2009, AJ, 137, 3718) and of
NASA's Astrophysics Data System.
\end{acknowledgements}
  
\bibliographystyle{raa}
\bibliography{references.bib}

\begin{thebibliography}{76}
\providecommand\natexlab[1]{#1}
\providecommand\JournalTitle[1]{#1}

\bibitem[{Abdo} {et~al.}(2009{\natexlab{a}})]{2009ApJ...707..727A}
{Abdo}, A.~A., {Ackermann}, M., {Ajello}, M., {et~al.} 2009{\natexlab{a}},
  \apj, 707, 727

\bibitem[{Abdo} {et~al.}(2009{\natexlab{b}})]{2009ApJ...707L.142A}
{Abdo}, A.~A., {Ackermann}, M., {Ajello}, M., {et~al.} 2009{\natexlab{b}},
  \apjl, 707, L142

\bibitem[{Angelakis} {et~al.}(2010)]{2010arXiv1006.5610A}
{Angelakis}, E., {Fuhrmann}, L., {Nestoras}, I., {et~al.} 2010, in {Proceedings
  of the Workshop Fermi meets Jansky - AGN in Radio and Gamma-Rays, MPIfR,
  Bonn}, ed. {Savolainen, T., Ros, E., Porcas, R.W. \& Zensus, J.A.}, {}

\bibitem[{Angelakis} {et~al.}(2009)]{2009A&A...501..801A}
{Angelakis}, E., {Kraus}, A., {Readhead}, A.~C.~S., {et~al.} 2009, \aap, 501,
  801

\bibitem[{Angelakis} {et~al.}(2015)]{2015A&A...575A..55A}
{Angelakis}, E., {Fuhrmann}, L., {Marchili}, N., {et~al.} 2015, \aap, 575, A55

\bibitem[{Ant{\'o}n} {et~al.}(2008)]{2008A&A...490..583A}
{Ant{\'o}n}, S., {Browne}, I.~W.~A., \& {March{\~a}}, M.~J. 2008, \aap, 490,
  583

\bibitem[{Boller} {et~al.}(1996)]{1996A&A...305...53B}
{Boller}, T., {Brandt}, W.~N., \& {Fink}, H. 1996, \aap, 305, 53

\bibitem[{Boroson} \& {Green}(1992)]{1992ApJS...80..109B}
{Boroson}, T.~A., \& {Green}, R.~F. 1992, \apjs, 80, 109

\bibitem[{Calderone} {et~al.}(2011)]{2011MNRAS.413.2365C}
{Calderone}, G., {Foschini}, L., {Ghisellini}, G., {et~al.} 2011, \mnras, 413,
  2365

\bibitem[{Calderone} {et~al.}(2013)]{2013MNRAS.431..210C}
{Calderone}, G., {Ghisellini}, G., {Colpi}, M., \& {Dotti}, M. 2013, \mnras,
  431, 210

\bibitem[{Carpenter} \& {Ojha}(2013)]{2013ATel.5344....1C}
{Carpenter}, B., \& {Ojha}, R. 2013, The Astronomer's Telegram, 5344

\bibitem[{Collin} {et~al.}(2006)]{2006A&A...456...75C}
{Collin}, S., {Kawaguchi}, T., {Peterson}, B.~M., \& {Vestergaard}, M. 2006,
  \aap, 456, 75

\bibitem[{D'Ammando} {et~al.}(2012)]{2012MNRAS.426..317D}
{D'Ammando}, F., {Orienti}, M., {Finke}, J., {et~al.} 2012, \mnras, 426, 317

\bibitem[{D'Ammando} {et~al.}(2013)]{2013MNRAS.436..191D}
{D'Ammando}, F., {Orienti}, M., {Finke}, J., {et~al.} 2013, \mnras, 436, 191

\bibitem[{Doi} {et~al.}(2011)]{2011ApJ...738..126D}
{Doi}, A., {Asada}, K., \& {Nagai}, H. 2011, \apj, 738, 126

\bibitem[{Doi} {et~al.}(2012)]{2012ApJ...760...41D}
{Doi}, A., {Nagira}, H., {Kawakatu}, N., {et~al.} 2012, \apj, 760, 41

\bibitem[{Foschini} {et~al.}(2009)]{2009AdSpR..43..889F}
{Foschini}, L., {Maraschi}, L., {Tavecchio}, F., {et~al.} 2009, Advances in
  Space Research, 43, 889

\bibitem[{Foschini} {et~al.}(2012)]{2012A&A...548A.106F}
{Foschini}, L., {Angelakis}, E., {Fuhrmann}, L., {et~al.} 2012, \aap, 548, A106

\bibitem[{Foschini} {et~al.}(2015)]{2015A&A...575A..13F}
{Foschini}, L., {Berton}, M., {Caccianiga}, A., {et~al.} 2015, \aap, 575, A13

\bibitem[{Fuhrmann}(2010)]{2010arXiv1007.0348F}
{Fuhrmann}, L. 2010, in {Proceedings of the Workshop Fermi meets Jansky - AGN
  in Radio and Gamma-Rays, MPIfR, Bonn}, ed. {Savolainen, T., Ros, E., Porcas,
  R.W. \& Zensus, J.A.}, {}

\bibitem[{Fuhrmann} {et~al.}(2007)]{2007AIPC..921..249F}
{Fuhrmann}, L., {Zensus}, J.~A., {Krichbaum}, T.~P., {Angelakis}, E., \&
  {Readhead}, A.~C.~S. 2007, in {American Institute of Physics Conference
  Series}, Vol. 921, {The First GLAST Symposium}, ed. {S.~Ritz, P.~Michelson,
  \& C.~A.~Meegan}, 249

\bibitem[{Fuhrmann} {et~al.}(2008)]{2008A&A...490.1019F}
{Fuhrmann}, L., {Krichbaum}, T.~P., {Witzel}, A., {et~al.} 2008, \aap, 490,
  1019

\bibitem[{Fuhrmann} {et~al.}(2011)]{2011nlsg.confE..26F}
{Fuhrmann}, L., {Angelakis}, E., {Nestoras}, I., {et~al.} 2011, in Narrow-Line
  Seyfert 1 Galaxies and their Place in the Universe, 26

\bibitem[{Fuhrmann} {et~al.}(2014)]{2014MNRAS.441.1899F}
{Fuhrmann}, L., {Larsson}, S., {Chiang}, J., {et~al.} 2014, \mnras, 441, 1899

\bibitem[{Grupe}(2004)]{2004AJ....127.1799G}
{Grupe}, D. 2004, \aj, 127, 1799

\bibitem[{Grupe} {et~al.}(2010)]{2010ApJS..187...64G}
{Grupe}, D., {Komossa}, S., {Leighly}, K.~M., \& {Page}, K.~L. 2010, \apjs,
  187, 64

\bibitem[{Gu} {et~al.}(2015)]{2015ApJS..221....3G}
{Gu}, M., {Chen}, Y., {Komossa}, S., {et~al.} 2015, \apjs, 221, 3

\bibitem[{H{\"o}gbom}(1974)]{1974A&AS...15..417H}
{H{\"o}gbom}, J.~A. 1974, \aaps, 15, 417

\bibitem[{Hovatta} {et~al.}(2009)]{2009A&A...494..527H}
{Hovatta}, T., {Valtaoja}, E., {Tornikoski}, M., \& {L{\"a}hteenm{\"a}ki}, A.
  2009, \aap, 494, 527

\bibitem[{Itoh} {et~al.}(2014)]{2014PASJ...66..108I}
{Itoh}, R., {Tanaka}, Y.~T., {Akitaya}, H., {et~al.} 2014, \pasj, 66, 108

\bibitem[{Jarvis} \& {McLure}(2006)]{2006MNRAS.369..182J}
{Jarvis}, M.~J., \& {McLure}, R.~J. 2006, \mnras, 369, 182

\bibitem[{Jorstad} {et~al.}(2005)]{2005AJ....130.1418J}
{Jorstad}, S.~G., {Marscher}, A.~P., {Lister}, M.~L., {et~al.} 2005, \aj, 130,
  1418

\bibitem[{Karamanavis}(2015)]{2015PhDT.......232K}
{Karamanavis}, V. 2015, {Zooming into gamma-ray loud galactic nuclei: broadband
  emission and structure dynamics of the blazar PKS 1502+106 and the
  narrow-line Seyfert 1 1H 0323+342}, PhD thesis, University of Cologne, 2015

\bibitem[{Karamanavis} {et~al.}(2016)]{2016A&A...586A..60K}
{Karamanavis}, V., {Fuhrmann}, L., {Krichbaum}, T.~P., {et~al.} 2016, \aap,
  586, A60

\bibitem[{Komossa}(2008)]{2008RMxAC..32...86K}
{Komossa}, S. 2008, in {}, Vol.~32, {Revista Mexicana de Astronomia y
  Astrofisica Conference Series}, 86

\bibitem[{Komossa} {et~al.}(2006)]{2006AJ....132..531K}
{Komossa}, S., {Voges}, W., {Xu}, D., {et~al.} 2006, \aj, 132, 531

\bibitem[{Komossa} {et~al.}(2008)]{2008ApJ...680..926K}
{Komossa}, S., {Xu}, D., {Zhou}, H., {Storchi-Bergmann}, T., \& {Binette}, L.
  2008, \apj, 680, 926

\bibitem[{Komossa} {et~al.}(2015)]{2015A&A...574A.121K}
{Komossa}, S., {Xu}, D., {Fuhrmann}, L., {et~al.} 2015, \aap, 574, A121

\bibitem[{Kreikenbohm} {et~al.}(2015)]{2015arXiv150903735K}
{Kreikenbohm}, A., {Schulz}, R., {Kadler}, M., {et~al.} 2015, arXiv:1509.03735

\bibitem[{Leighly}(1999)]{1999ApJS..125..297L}
{Leighly}, K.~M. 1999, \apjs, 125, 297

\bibitem[{Le{\'o}n Tavares} {et~al.}(2014)]{2014ApJ...795...58L}
{Le{\'o}n Tavares}, J., {Kotilainen}, J., {Chavushyan}, V., {et~al.} 2014,
  \apj, 795, 58

\bibitem[{Liao} {et~al.}(2015)]{2015arXiv151005584L}
{Liao}, N.-H., {Liang}, Y.-F., {Weng}, S.-S., {Gu}, M.-F., \& {Fan}, Y.-Z.
  2015, arXiv:1510.05584

\bibitem[{Lister} {et~al.}(2009)]{2009AJ....137.3718L}
{Lister}, M.~L., {Aller}, H.~D., {Aller}, M.~F., {et~al.} 2009, \aj, 137, 3718

\bibitem[{Lister} {et~al.}(2013)]{2013AJ....146..120L}
{Lister}, M.~L., {Aller}, M.~F., {Aller}, H.~D., {et~al.} 2013, \aj, 146, 120

\bibitem[{Mathur} {et~al.}(2012)]{2012ApJ...754..146M}
{Mathur}, S., {Fields}, D., {Peterson}, B.~M., \& {Grupe}, D. 2012, \apj, 754,
  146

\bibitem[{Neumann} {et~al.}(1994)]{1994A&AS..106..303N}
{Neumann}, M., {Reich}, W., {Fuerst}, E., {et~al.} 1994, \aaps, 106, 303

\bibitem[{Onuchukwu} \& {Ubachukwu}(2013)]{2013Ap&SS.348..193O}
{Onuchukwu}, C.~C., \& {Ubachukwu}, A.~A. 2013, \apss, 348, 193

\bibitem[{Orban de Xivry} {et~al.}(2011)]{2011MNRAS.417.2721O}
{Orban de Xivry}, G., {Davies}, R., {Schartmann}, M., {et~al.} 2011, \mnras,
  417, 2721

\bibitem[{Orienti} {et~al.}(2015)]{2015MNRAS.453.4037O}
{Orienti}, M., {D'Ammando}, F., {Larsson}, J., {et~al.} 2015, \mnras, 453, 4037

\bibitem[{Osterbrock} \& {Pogge}(1985)]{1985ApJ..297...166}
{Osterbrock}, D.~E., \& {Pogge}, R. 1985, \apj, 297, 166

\bibitem[{Paliya} {et~al.}(2014)]{2014ApJ...789..143P}
{Paliya}, V.~S., {Sahayanathan}, S., {Parker}, M.~L., {et~al.} 2014, \apj, 789,
  143

\bibitem[{Paliya} {et~al.}(2015)]{2015AJ....149...41P}
{Paliya}, V.~S., {Stalin}, C.~S., \& {Ravikumar}, C.~D. 2015, \aj, 149, 41

\bibitem[{Peterson}(2011)]{2011arXiv1109.4181P}
{Peterson}, B.~M. 2011, in Narrow-Line Seyfert 1 Galaxies and their Place in
  the Universe

\bibitem[{Puchnarewicz} {et~al.}(1992)]{1992MNRAS.256..589P}
{Puchnarewicz}, E.~M., {Mason}, K.~O., {Cordova}, F.~A., {et~al.} 1992, \mnras,
  256, 589

\bibitem[{Readhead}(1994)]{Readhead1994ApJ}
{Readhead}, A.~C.~S. 1994, \apj, 426, 51

\bibitem[{Rees}(1966)]{1966Natur.211..468R}
{Rees}, M.~J. 1966, \nat, 211, 468

\bibitem[{Richards} \& {Lister}(2015)]{2015ApJ...800L...8R}
{Richards}, J.~L., \& {Lister}, M.~L. 2015, \apjl, 800, L8

\bibitem[{Savolainen} {et~al.}(2010)]{2010A&A...512A..24S}
{Savolainen}, T., {Homan}, D.~C., {Hovatta}, T., {et~al.} 2010, \aap, 512, A24

\bibitem[{Scheuer} \& {Readhead}(1979)]{1979Natur.277..182S}
{Scheuer}, P.~A.~G., \& {Readhead}, A.~C.~S. 1979, \nat, 277, 182

\bibitem[{Schulz} {et~al.}(2015)]{2015arXiv151102631S}
{Schulz}, R., {Kreikenbohm}, A., {Kadler}, M., {et~al.} 2015, arXiv:1511.02631

\bibitem[{Shepherd} {et~al.}(1994)]{1994BAAS...26..987S}
{Shepherd}, M.~C., {Pearson}, T.~J., \& {Taylor}, G.~B. 1994, in Bulletin of
  the American Astronomical Society, Vol.~26, Bulletin of the American
  Astronomical Society, 987

\bibitem[{Smith} {et~al.}(2005)]{2005MNRAS.359..846S}
{Smith}, J.~E., {Robinson}, A., {Young}, S., {Axon}, D.~J., \& {Corbett}, E.~A.
  2005, \mnras, 359, 846

\bibitem[{Spergel} {et~al.}(2007)]{2007ApJS..170..377S}
{Spergel}, D.~N., {Bean}, R., {Dor{\'e}}, O., {et~al.} 2007, \apjs, 170, 377

\bibitem[{Sulentic} {et~al.}(2000)]{2000ApJ...536L...5S}
{Sulentic}, J.~W., {Zwitter}, T., {Marziani}, P., \& {Dultzin-Hacyan}, D. 2000,
  \apjl, 536, L5

\bibitem[{Sun} {et~al.}(2015)]{2015ApJ...798...43S}
{Sun}, X.-N., {Zhang}, J., {Lin}, D.-B., {et~al.} 2015, \apj, 798, 43

\bibitem[{Ungerechts} {et~al.}(1998)]{1998ASPC..144..149U}
{Ungerechts}, H., {Kramer}, C., {Lefloch}, B., {et~al.} 1998, in {Astronomical
  Society of the Pacific Conference Series}, Vol. 144, {IAU Colloq. 164: Radio
  Emission from Galactic and Extragalactic Compact Sources}, ed. {J.~A.~Zensus,
  G.~B.~Taylor, \& J.~M.~Wrobel}, 149

\bibitem[{V{\'e}ron-Cetty} {et~al.}(2001)]{2001A&A...372..730V}
{V{\'e}ron-Cetty}, M.-P., {V{\'e}ron}, P., \& {Gon{\c c}alves}, A.~C. 2001,
  \aap, 372, 730

\bibitem[{Wajima} {et~al.}(2014)]{2014ApJ...781...75W}
{Wajima}, K., {Fujisawa}, K., {Hayashida}, M., {et~al.} 2014, \apj, 781, 75

\bibitem[{Xu} {et~al.}(2012)]{2012AJ....143...83X}
{Xu}, D., {Komossa}, S., {Zhou}, H., {et~al.} 2012, \aj, 143, 83

\bibitem[{Yao} {et~al.}(2015{\natexlab{a}})]{2015AJ....150...23Y}
{Yao}, S., {Yuan}, W., {Komossa}, S., {et~al.} 2015{\natexlab{a}}, \aj, 150, 23

\bibitem[{Yao} {et~al.}(2015{\natexlab{b}})]{2015MNRAS.454L..16Y}
{Yao}, S., {Yuan}, W., {Zhou}, H., {et~al.} 2015{\natexlab{b}}, \mnras, 454,
  L16

\bibitem[{Zhang} {et~al.}(2013)]{2013ApJ...774L...5Z}
{Zhang}, J., {Liang}, E.-W., {Sun}, X.-N., {et~al.} 2013, \apjl, 774, L5

\bibitem[{Zhang} {et~al.}(2012)]{2012ApJ...752..157Z}
{Zhang}, J., {Liang}, E.-W., {Zhang}, S.-N., \& {Bai}, J.~M. 2012, \apj, 752,
  157

\bibitem[{Zhang} {et~al.}(2014)]{2014ApJ...788..104Z}
{Zhang}, J., {Sun}, X.-N., {Liang}, E.-W., {et~al.} 2014, \apj, 788, 104

\bibitem[{Zhou} {et~al.}(2006)]{2006ApJS..166..128Z}
{Zhou}, H., {Wang}, T., {Yuan}, W., {et~al.} 2006, \apjs, 166, 128

\bibitem[{Zhou} {et~al.}(2007)]{2007ApJ...658L..13Z}
{Zhou}, H., {Wang}, T., {Yuan}, W., {et~al.} 2007, \apjl, 658, L13

\end{thebibliography}


\appendix\label{appendix}
\section{MODELFIT Results}
\begin{table*}
  \caption{\textsc{MODELFIT} results of the VLBI imaging at 15\,GHz for
1H\,0323+342. Columns from     left to right: (1) observing epoch in fractional
year, (2)     integrated component flux density, (3) radial separation from the
    core, (4) position angle, (5) component size given as the FWHM of
    the major axis, (6) component identification label.  C denotes the
    core, S the quasi-stationary component and C\# the rest of the
    moving features. The designation `un' refers to non-robustly
cross-identified features.}
\centering
\begin{tabular}{c c c c c c}
\hline
Epoch    &    $S$     &   $r$    &      PA      &  FWHM   &   ID\\
         &   (Jy)     & (mas)  & ($^{\circ}$) &  (mas)  &     \\
\hline
\noalign{\smallskip}
2010.79  &  $0.193 \pm 0.019$   &        \dots        &         \dots        &  

$ 0.08 \pm 0.01$ &  C	\\
2010.79  &  $0.064 \pm 0.006$   &  $ 0.29 \pm 0.03 $  &  $ 131.6 \pm 6.2  $  &  

$ 0.16 \pm 0.02$ &  S	\\
2010.79  &  $0.019 \pm 0.002$   &  $ 0.96 \pm 0.08 $  &  $ 127.0 \pm 4.8  $  &  

$ 0.41 \pm 0.04$ &  C5  \\
2010.79  &  $0.011 \pm 0.001$   &  $ 1.69 \pm 0.11 $  &  $ 123.9 \pm 3.6  $  &  

$ 0.53 \pm 0.05$ &  C4  \\
2010.79  &  $0.002 \pm 0.001$   &  $ 3.11 \pm 0.19 $  &  $ 123.4 \pm 3.5  $  &  

$ 0.94 \pm 0.09$ &  un  \\
2010.79  &  $0.004 \pm 0.001$   &  $ 5.82 \pm 0.13 $  &  $ 123.5 \pm 1.3  $  &  

$ 0.67 \pm 0.07$ &  C2  \\
2010.79  &  $0.011 \pm 0.001$   &  $ 7.21 \pm 0.17 $  &  $ 125.2 \pm 1.4  $  &  

$ 0.87 \pm 0.09$ &  C1  \\
2010.79  &  $0.005 \pm 0.001$   &  $ 8.81 \pm 0.24 $  &  $ 125.9 \pm 1.6  $  &  

$ 1.21 \pm 0.12$ &  C0  \\
\noalign{\smallskip}
2011.17  &  $0.322 \pm 0.032$   &	 \dots          & 	  \dots	     &  

$ 0.10 \pm 0.01$ &  C	\\
2011.17  &  $0.048 \pm 0.005$   &  $ 0.28 \pm 0.02 $  &  $ 131.6 \pm 4.0  $  &  

$ 0.10 \pm 0.01$ &  S	\\
2011.17  &  $0.012 \pm 0.001$   &  $ 0.93 \pm 0.06 $  &  $ 124.2 \pm 3.8  $  &  

$ 0.31 \pm 0.03$ &  C5  \\
2011.17  &  $0.014 \pm 0.001$   &  $ 1.94 \pm 0.13 $  &  $ 124.4 \pm 3.8  $  &  

$ 0.65 \pm 0.07$ &  C4  \\
2011.17  &  $0.003 \pm 0.001$   &  $ 5.39 \pm 0.16 $  &  $ 122.9 \pm 1.7  $  &  

$ 0.82 \pm 0.08$ &  C3  \\
2011.17  &  $0.009 \pm 0.001$   &  $ 6.85 \pm 0.13 $  &  $ 124.5 \pm 1.1  $  &  

$ 0.66 \pm 0.07$ &  C2  \\
2011.17  &  $0.008 \pm 0.001$   &  $ 7.95 \pm 0.18 $  &  $ 125.3 \pm 1.3  $  &  

$ 0.89 \pm 0.09$ &  C1  \\
\noalign{\smallskip}
2011.40  &  $0.251 \pm 0.025$   &	 \dots          & 	  \dots	     &  

$ 0.10 \pm 0.01$ &  C	\\
2011.40  &  $0.048 \pm 0.005$   &  $ 0.34 \pm 0.02 $  &  $ 135.1 \pm 3.2  $  &  

$ 0.10 \pm 0.01$ &  S	\\
2011.40  &  $0.010 \pm 0.001$   &  $ 0.68 \pm 0.02 $  &  $ 127.5 \pm 1.6  $  &  

$ 0.10 \pm 0.01$ &  C6  \\
2011.40  &  $0.006 \pm 0.001$   &  $ 1.38 \pm 0.08 $  &  $ 126.7 \pm 3.4  $  &  

$ 0.41 \pm 0.04$ &  C5  \\
2011.40  &  $0.008 \pm 0.001$   &  $ 2.45 \pm 0.17 $  &  $ 124.5 \pm 4.0  $  &  

$ 0.85 \pm 0.09$ &  C4  \\
2011.40  &  $0.006 \pm 0.001$   &  $ 6.15 \pm 0.18 $  &  $ 124.6 \pm 1.7  $  &  

$ 0.90 \pm 0.09$ &  C3  \\
2011.40  &  $0.008 \pm 0.001$   &  $ 7.06 \pm 0.11 $  &  $ 125.0 \pm 0.9  $  &  

$ 0.56 \pm 0.06$ &  C2  \\
2011.40  &  $0.005 \pm 0.001$   &  $ 8.09 \pm 0.16 $  &  $ 125.2 \pm 1.1  $  &  

$ 0.78 \pm 0.08$ &  C1  \\
\noalign{\smallskip}
2011.53  &  $0.194 \pm 0.019$   &	 \dots          & 	  \dots	     &  

$ 0.10 \pm 0.01$ &  C	\\
2011.53  &  $0.053 \pm 0.005$   &  $ 0.26 \pm 0.02 $  &  $ 133.4 \pm 4.2  $  &  

$ 0.10 \pm 0.01$ &  S	\\
2011.53  &  $0.013 \pm 0.001$   &  $ 0.70 \pm 0.02 $  &  $ 129.4 \pm 1.6  $  &  

$ 0.10 \pm 0.01$ &  C6  \\
2011.53  &  $0.006 \pm 0.001$   &  $ 1.45 \pm 0.08 $  &  $ 124.8 \pm 3.1  $  &  

$ 0.39 \pm 0.04$ &  C5  \\
2011.53  &  $0.007 \pm 0.001$   &  $ 2.69 \pm 0.17 $  &  $ 125.8 \pm 3.6  $  &  

$ 0.86 \pm 0.09$ &  C4  \\
2011.53  &  $0.005 \pm 0.001$   &  $ 6.27 \pm 0.19 $  &  $ 123.4 \pm 1.7  $  &  

$ 0.96 \pm 0.10$ &  C3  \\
2011.53  &  $0.010 \pm 0.001$   &  $ 7.19 \pm 0.11 $  &  $ 124.9 \pm 0.9  $  &  

$ 0.55 \pm 0.06$ &  C2  \\
2011.53  &  $0.007 \pm 0.001$   &  $ 8.99 \pm 0.37 $  &  $ 125.0 \pm 2.3  $  &  

$ 1.83 \pm 0.18$ &  C1  \\
\noalign{\smallskip}
2011.99  &  $0.314 \pm 0.031$   &	 \dots          & 	  \dots	     &  

$ 0.10 \pm 0.01$ &  C	\\
2011.99  &  $0.021 \pm 0.002$   &  $ 0.34 \pm 0.02 $  &  $ 129.6 \pm 3.2  $  &  

$ 0.10 \pm 0.01$ &  S	\\
2011.99  &  $0.006 \pm 0.001$   &  $ 0.89 \pm 0.02 $  &  $ 129.7 \pm 1.2  $  &  

$ 0.10 \pm 0.01$ &  C6  \\
2011.99  &  $0.008 \pm 0.001$   &  $ 1.44 \pm 0.10 $  &  $ 130.9 \pm 3.9  $  &  

$ 0.49 \pm 0.05$ &  C5  \\
2011.99  &  $0.008 \pm 0.001$   &  $ 3.53 \pm 0.34 $  &  $ 121.3 \pm 5.5  $  &  

$ 1.71 \pm 0.17$ &  C4  \\
2011.99  &  $0.009 \pm 0.001$   &  $ 6.66 \pm 0.21 $  &  $ 124.5 \pm 1.8  $  &  

$ 1.03 \pm 0.10$ &  C3  \\
2011.99  &  $0.006 \pm 0.001$   &  $ 7.52 \pm 0.07 $  &  $ 125.0 \pm 0.6  $  &  

$ 0.37 \pm 0.04$ &  C2  \\
2011.99  &  $0.006 \pm 0.001$   &  $ 8.29 \pm 0.15 $  &  $ 124.2 \pm 1.0  $  &  

$ 0.75 \pm 0.08$ &  C1  \\
\noalign{\smallskip}
\hline
\end{tabular}
\label{tab:modelfit}
\end{table*}
%
%
\begin{table*}
\centering
\begin{tabular}{c c c c c c}
\hline
Epoch    &    S     &	r    &      PA      &  FWHM   &   ID\\
         &   (Jy)   & (mas)  & ($^{\circ}$) &  (mas)  &     \\
\hline
\noalign{\smallskip}
2012.53  &  $0.181 \pm 0.018$   &	 \dots          & 	  \dots	     &  

$ 0.10 \pm 0.01$ &  C	\\
2012.53  &  $0.029 \pm 0.003$   &  $ 0.30 \pm 0.02 $  &  $ 132.8 \pm 3.7  $  &  

$ 0.10 \pm 0.01$ &  S	\\
2012.53  &  $0.004 \pm 0.001$   &  $ 1.02 \pm 0.05 $  &  $ 128.9 \pm 2.5  $  &  

$ 0.23 \pm 0.02$ &  C6  \\
2012.53  &  $0.005 \pm 0.001$   &  $ 2.34 \pm 0.16 $  &  $ 125.1 \pm 3.8  $  &  

$ 0.78 \pm 0.08$ &  C5  \\
2012.53  &  $0.006 \pm 0.001$   &  $ 4.67 \pm 0.29 $  &  $ 124.0 \pm 3.6  $  &  

$ 1.46 \pm 0.15$ &  C4  \\
2012.53  &  $0.004 \pm 0.001$   &  $ 7.04 \pm 0.03 $  &  $ 125.2 \pm 0.2  $  &  

$ 0.15 \pm 0.02$ &  C3  \\
2012.53  &  $0.005 \pm 0.001$   &  $ 7.69 \pm 0.10 $  &  $ 124.6 \pm 0.7  $  &  

$ 0.50 \pm 0.05$ &  C2  \\
2012.53  &  $0.004 \pm 0.001$   &  $ 8.89 \pm 0.19 $  &  $ 125.2 \pm 1.2  $  &  

$ 0.96 \pm 0.10$ &  C1  \\
\noalign{\smallskip}
2012.98  &  $0.194 \pm 0.019$   &	 \dots          & 	  \dots	     &  

$ 0.10 \pm 0.01$ &  C	\\
2012.98  &  $0.040 \pm 0.004$   &  $ 0.33 \pm 0.02 $  &  $ 125.5 \pm 3.3  $  &  

$ 0.10 \pm 0.01$ &  S	\\
2012.98  &  $0.006 \pm 0.001$   &  $ 0.88 \pm 0.04 $  &  $ 133.0 \pm 2.9  $  &  

$ 0.22 \pm 0.02$ &  C6  \\
2012.98  &  $0.004 \pm 0.001$   &  $ 2.45 \pm 0.31 $  &  $ 124.3 \pm 7.1  $  &  

$ 1.53 \pm 0.15$ &  C5  \\
2012.98  &  $0.009 \pm 0.001$   &  $ 5.03 \pm 0.38 $  &  $ 123.5 \pm 4.4  $  &  

$ 1.92 \pm 0.19$ &  C4  \\
2012.98  &  $0.003 \pm 0.001$   &  $ 6.56 \pm 0.08 $  &  $ 133.4 \pm 0.7  $  &  

$ 0.41 \pm 0.04$ &  C3  \\
2012.98  &  $0.011 \pm 0.001$   &  $ 7.88 \pm 0.37 $  &  $ 124.9 \pm 2.7  $  &  

$ 1.85 \pm 0.19$ &  C2  \\
\noalign{\smallskip}
2013.52  &  $0.139 \pm 0.014$   &	 \dots          & 	  \dots	     &  

$ 0.10 \pm 0.01$ &  C	\\
2013.52  &  $0.031 \pm 0.003$   &  $ 0.36 \pm 0.02 $  &  $ 128.5 \pm 3.1  $  &  

$ 0.10 \pm 0.01$ &  S	\\
2013.52  &  $0.019 \pm 0.002$   &  $ 0.67 \pm 0.02 $  &  $ 126.3 \pm 1.6  $  &  

$ 0.10 \pm 0.01$ &  NC  \\
2013.52  &  $0.003 \pm 0.001$   &  $ 1.26 \pm 0.08 $  &  $ 132.4 \pm 3.7  $  &  

$ 0.41 \pm 0.04$ &  C6  \\
2013.52  &  $0.002 \pm 0.001$   &  $ 2.15 \pm 0.02 $  &  $ 124.5 \pm 0.5  $  &  

$ 0.10 \pm 0.01$ &  un  \\
2013.52  &  $0.003 \pm 0.001$   &  $ 4.53 \pm 0.21 $  &  $ 129.3 \pm 2.6  $  &  

$ 1.03 \pm 0.10$ &  C5  \\
2013.52  &  $0.008 \pm 0.001$   &  $ 6.19 \pm 0.16 $  &  $ 124.3 \pm 1.5  $  &  

$ 0.79 \pm 0.08$ &  C4  \\
2013.52  &  $0.008 \pm 0.001$   &  $ 7.07 \pm 0.08 $  &  $ 124.7 \pm 0.6  $  &  

$ 0.39 \pm 0.04$ &  C3  \\
2013.52  &  $0.004 \pm 0.001$   &  $ 8.10 \pm 0.17 $  &  $ 124.5 \pm 1.2  $  &  

$ 0.84 \pm 0.08$ &  C2  \\
2013.52  &  $0.002 \pm 0.001$   &  $10.28 \pm 0.08 $  &  $ 127.4 \pm 0.4  $  &  

$ 0.38 \pm 0.04$ &  C1  \\
\noalign{\smallskip}
\hline
\end{tabular}
\end{table*}

\end{document}